\documentclass[noshowpacs,twocolumn,aps,pre]{revtex4}
\usepackage[dvips]{graphicx}
\usepackage{amssymb}
\usepackage{amsmath}
\usepackage{latexsym}
\usepackage{epsfig}
\usepackage{bm}
\usepackage{times,psfrag,subfigure}        
\usepackage{float}
\usepackage{graphicx,amsmath,amssymb}

\newcommand{\theeq}{\theta_\mathrm{eq}}
\newcommand{\thetr}{\theta_{\mathrm{tr}}}
\newcommand{\thev}{\theta_{\mathrm{v}}}

\begin{document}

\title{Using electrowetting to control interface motion in 
patterned microchannels}

\author{B.\ M.\ Mognetti}
\author{J.\ M.\ Yeomans}

\affiliation{
The Rudolf Peierls Centre for Theoretical Physics,
 1 Keble Road Oxford, OX1 3NP United Kindom. 
}

\date{\today}

\begin{abstract}
We use mesoscale simulations to demonstrate the feasibility of a novel 
microfluidic valve, which exploits Gibbs' pinning in microchannels patterned 
by posts or ridges, together with electrowetting.
\end{abstract}

\maketitle

As fluidic devices become smaller there is a need for simple and robust ways 
to control the motion of fluids moving through them. Several way to arrest 
capillary filling have been developed  
\cite{CH-04,SYHOS-08,MS-06,BB-98,HFE-99,DGLSK-99}. These are mainly based 
on using hydrophobic patches  \cite{CH-04,SYHOS-08,MS-06} or narrow hydrophobic
 channels to arrest the flow \cite{BB-98,HFE-99,DGLSK-99}. In the first case 
filling is restarted by an electric field which decreases the contact angle, 
while in the second a driving pressure is applied. In this paper we use 
numerical simulations to motivate an alternative mechanism for creating a 
microfluidic valve. We show that placing obstacles at relevant positions 
along a microchannel, and addressing them by an electrowetting potential 
applied at the position of the interface  \cite{QB-01}, can allow capillary 
filling \cite{LW} to be halted and restarted reversibly.

If a microchannel is brought into contact with a fluid, and the contact angle 
between the surface and the fluid, $\theeq$, is less than 90$^\circ$, the fluid 
will move into the channel \cite{LW}.  This is the well-known 
process of capillary filling. However, if there are obstacles to the flow, 
such as micron-scale ridges or posts on the surface, the advancing front 
can pin for contact angles greater than a certain threshold value 
($\thetr< 90^\circ$), thus halting the filling. 
For a slowly moving interface
the pinning can be understood in terms of the Gibbs' criterion \cite{Gibbs} which
states that an interface remains pinned to a corner as long as it makes
an angle smaller than $\theeq$ with each of the dry 
surfaces meeting at the corner. As the interface is pulled along the channel 
by the capillary forces, it may not reach the required angle to 
wet the surface of the obstacles, and hence the filling is arrested.

\begin{figure}[!h]
 \includegraphics[angle=0,scale=0.25]{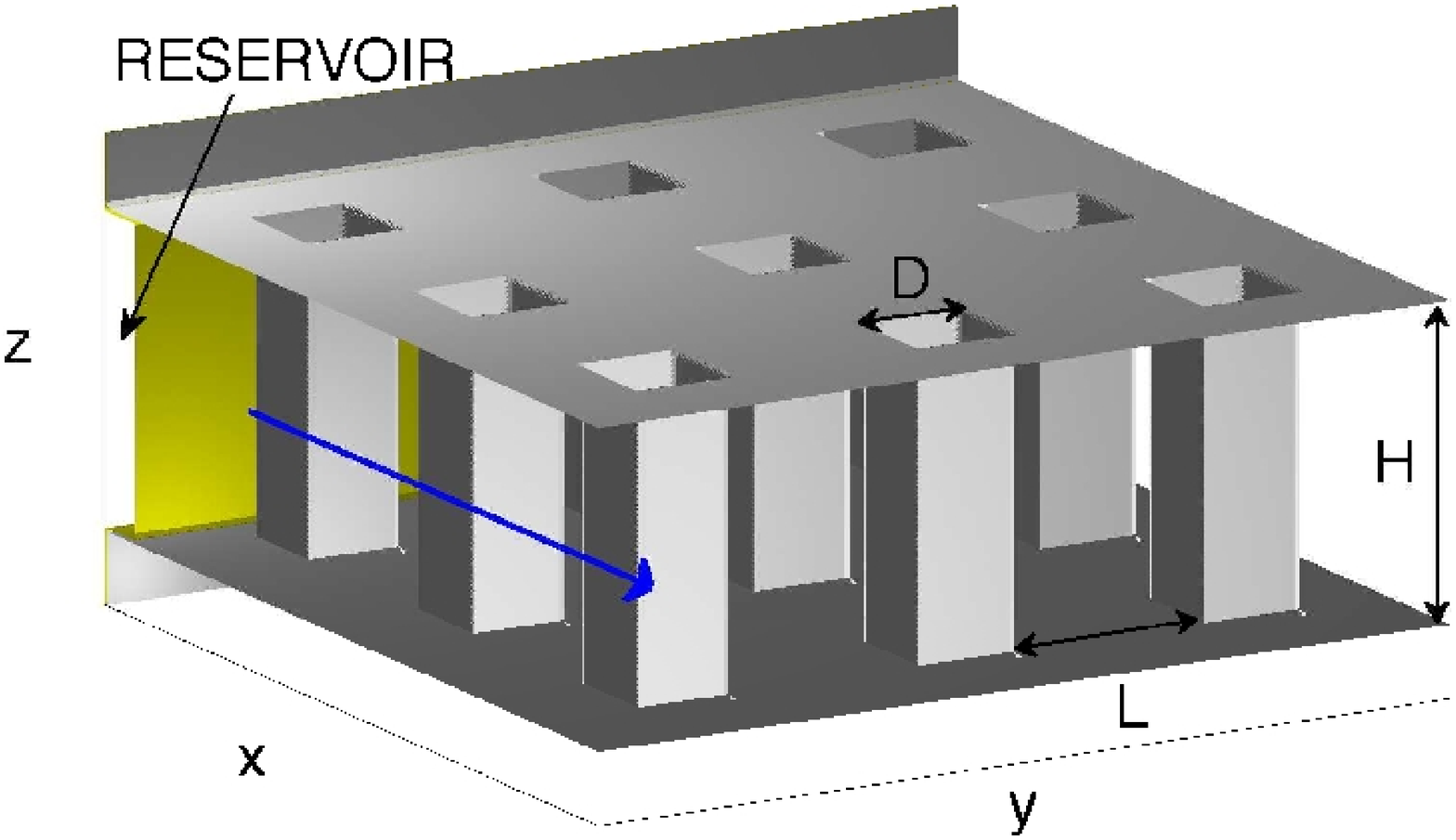} (a) \\ \vspace{0.7cm}
 \includegraphics[angle=0,scale=0.41]{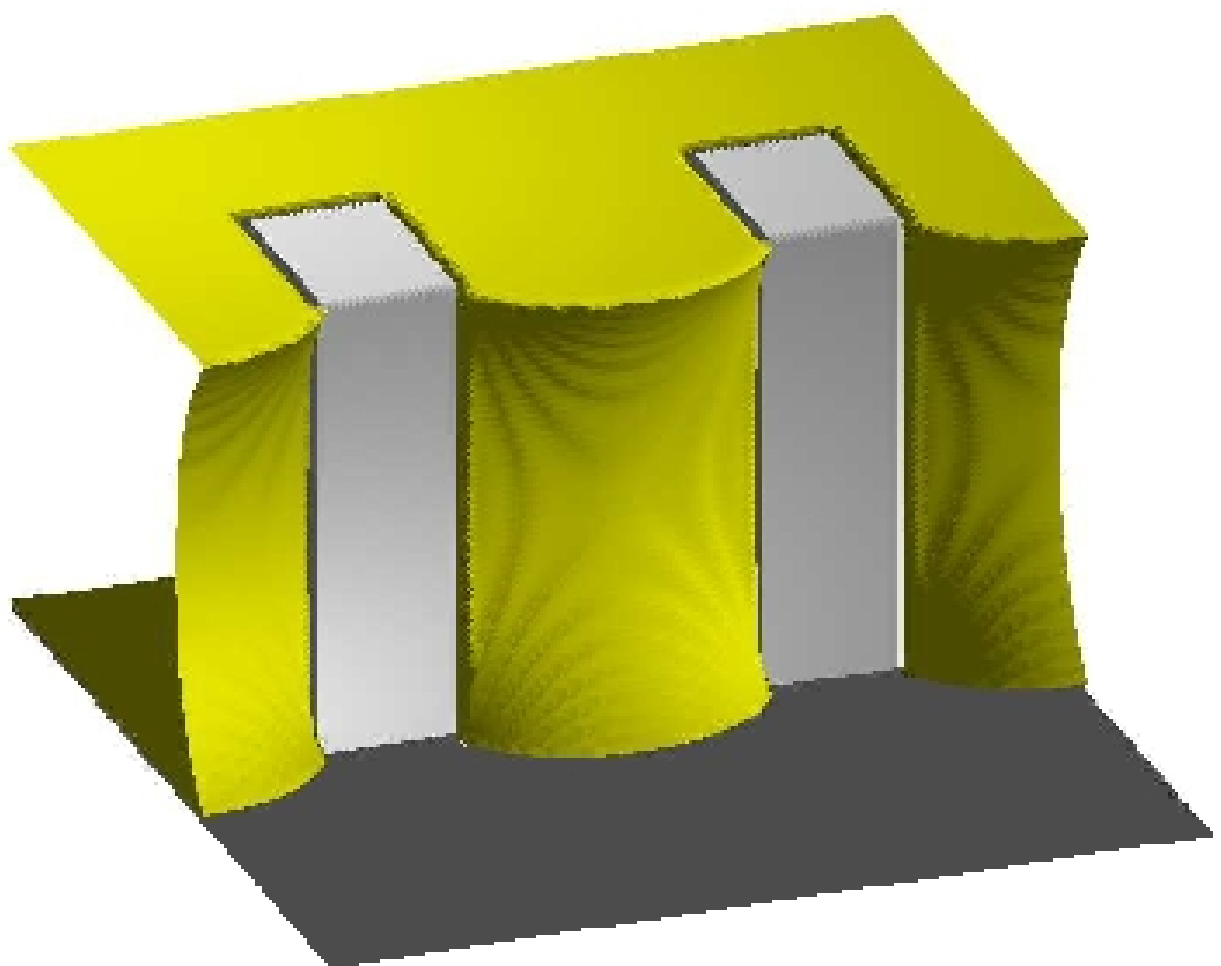} (b) \\ \vspace{0.7cm}  
 \includegraphics[angle=0,scale=0.25]{Fig1c.ps} (c)
\caption{
(a) Channel geometry: the fluid
flows along the $x$ direction and periodic boundary conditions are used
along the $y$ direction. (b) Typical shape of the front when it is pinned at a row of posts. 
For clarity the top wall is omitted. (c)  Position of the filling fluid  versus 
time. In these simulations $\theeq=60^\circ$, $\thev=30^\circ$ and $\thetr \sim 55^\circ$. 
}\label{fig1}
\end{figure}

We consider first the geometry shown in Fig.~\ref{fig1}{\em a}.
Posts of square cross section $D \times D$ and spacing $L$ reach across a 
channel of height $H$. Flow is along the $x$ direction and we consider a channel sufficiently wide that the 
effect of the side walls may be neglected. 
Fig.\ \ref{fig1}{\em b} depicts a
pinned configuration for this geometry. Although the front is pulled along the top and bottom walls of
the channel, for sufficiently large contact angles it does not move
far enough to start wetting the dry face of the posts.
For $H/L>1$ the threshold contact angle below which flow is possible is 
$\thetr \sim 55^\circ$. This value is independent of the channel height because the walls act independently 
\cite{MY-09}. As $H/L$ decreases below unity, $\thetr$ increases to $90^\circ$ as $H/L\to 0$. Hence a 
suitable choice of $H/L$ allows $\thetr$ to be tuned as required.

Electrowetting \cite{QB-01}, the application of an electric potential to a 
microchannel, is a robust method of changing the contact angles by
tens of degrees. Hence one might expect that the application of an
electrowetting potential can rather easily move the system between
the pinned and the filling regimes, thus allowing control of the motion of 
the capillary front. In this letter we present simulations of capillary 
filling in a patterned microchannel showing, in particular, how switching between contact angles 
above and below $\thetr$ can be used to halt and restart 
capillary filling. Before presenting our results we first outline the numerical approach we use.

As we are considering micron length scales it is appropriate to describe 
the system using a mesoscale algorithm. We consider
a binary fluid with components $A$ and $B$, say, which in equilibrium
is described by minimising the free energy functional
\begin{eqnarray}
\Psi = \int_\Omega \Big[ 
{c^2 \over 3} n \log n + {\kappa\over 2}(\partial_\alpha \phi)^2- {a\over 2}
\phi^2 +{a\over 4} \phi^4 \Big]  +\int_{\partial \Omega} h\cdot \phi \, ,
\nonumber \\ 
\label{eq1}
\end{eqnarray}
where $n$ is the local total density of the $A$ and $B$ components 
($n = n_A +n_B$), $\phi$ is the order parameter $\phi = n_A - n_B$ and $c$ 
is the lattice velocity $c = \delta x / \delta t$,  where
$\delta x$ is the lattice spacing and $\delta t$ is the simulation time step. 
The first integral in Eq.\ (\ref{eq1}), taken over the total volume $\Omega$, 
 controls the bulk properties of the system. The terms in $\phi$ give 
coexistence of phases with $\phi =  \pm 1$. The energy cost of an interface 
between the two phases is modeled by the derivative term, with ${\kappa}$ 
related to the surface tension. The term in $n$ controls the compressibility 
of the fluid.
The integral over the solid-fluid interface $\partial \Omega$ 
in Eq.\ (\ref{eq1}) accounts for the wetting properties of the solid surfaces
\cite{C-77}. 
The surface field $h$ can be used to tune the equilibrium contact angle.

The hydrodynamics of the fluid is described by the Navier-Stokes 
equations for the density $\rho$ and the velocity field ${\bf v}$ together 
with a convection-diffusion equation for the binary order parameter $\phi$
\begin{eqnarray}
& \partial_t \rho + \nabla \cdot (\rho \, {\bf v}) = 0 \, , & 
\label{eq2}\\
& \partial_t (\rho\, {\bf v}) + {\bf v} \cdot \nabla (\rho \, {\bf v}) =
-\nabla\cdot \underline{\underline{P}} + \eta \nabla^2  {\bf v}  \, , & 
\label{eq3}\\
& \partial_t \phi + \nabla \cdot (\phi \, {\bf v}) = M \nabla^2 \mu \, . &
\label{eq4}
\end{eqnarray}
In Eq.\ (\ref{eq3}) $\eta$ is the viscosity of the fluid and in Eq.\ 
(\ref{eq4}) $M$ is a mobility coefficient. The pressure tensor 
$\underline{\underline{P}}$
 and the chemical potential $\mu$ which appear in Eqs.\ (\ref{eq3}) and 
(\ref{eq4}), are calculated from the free energy (\ref{eq1}).

We have chosen to use a two-component binary fluid as a model system. 
This is because modeling capillary filling correctly using a liquid-gas 
diffuse interface model is computationally very demanding because of 
unphysical motion of the interface due to evaporation-condensation 
effects \cite{EC}. However our
results are equally applicable to a physical system where a liquid
displaces a gas as the important physical parameters are the viscosities, 
not the densities, of the fluid components. Therefore we shall
use the natural terminology `liquid' and `gas' for the displacing and
displaced fluid from now on.
The equations of motion are solved using a lattice Boltzmann
algorithm \cite{LB}. 
Full details are given in Refs.\ \cite{detail,Halim-08}.
In particular \cite{Halim-08}  shows that this
approach gives excellent agreement with analytic results for capillary
filling in smooth microchannels.

For our first simulations we investigate the microchannel geometry shown in Fig.~\ref{fig1}{\em a}. Measuring 
distances in lattice units, the simulation box is of size $(L_x;L_y;L_z) = (180; 60; 60)$ and we take 
periodic boundary conditions in the y-direction. The height of the channel
$H=60$, the separation of the posts $L=40$, and the posts are of width 
$D=20$ and lie between $x=20$ and $x=40$, $x=80$ and $x=100$, etc.
At the beginning and end of the channel are liquid and gas reservoirs respectively, connected
in order to equalise the pressure. The simulations commence with the interface at $x=0$ and with the 
elctrowetting potential switched off.

The subsequent position of the interface (recorded in the centre of channel)
as a function of time is shown in Fig.\ \ref{fig1}{\em c}. 
The zero potential equilibrium contact angle of the channel is chosen 
to be  $\theeq = 60^\circ$. This is 
greater than $\thetr\approx 55^\circ$ so the advancing front
is pinned at the first row of posts, in the configuration reported in
Fig.\ \ref{fig1}{\em b}. At $t = 2 \cdot 10^5 \,\delta t$ the  
contact angle of the channel is decreased to $\thev = 30^\circ$, 
corresponding to the application of an electrowetting field. Filling restarts
immediately and the interface advances. It is accelerated (by the
increased hydrophilic surface area) as it passes the second row of
posts and then slowed as it reaches the trailing edges of the posts, 
but is not pinned \cite{MY-09}. We have checked
that it also passes the third row of posts if $\thev$ is held at 30$^\circ$.

However, in the results presented in Fig.\ \ref{fig1}{\em c}, the 
equilibrium contact angle is switched back to $\theeq = 60^\circ$ 
when the advancing front reaches the middle point between the second
and the third posts.  This leads to pinning, explicitly showing the
possibility of stopping the imbibition of the channel at a prespecified
position.

As a more stringent test we also simulated a channel with six rows of posts
switching between $\theeq=60^\circ$ and $\thev=50^\circ$. Pinning and subsequent depinning when the 
electrowetting potential was activated was observed at each post.

The broken line in Fig.\ \ref{fig1}{\em c} compares a similar 
simulation, but with
the contact angle on the channel walls switched to $\thev=30^\circ$ but that on
the surfaces of the posts remaining at $\theeq=60^\circ$. The behaviour of the
fluid front is essentially the same, except for a slight slowing due to
the reduction in the capillary force. This demonstrates that the valve 
mechanism does not have a strong dependence on the details of the 
implementation of the electrowetting potential.
Moreover the applied electric field does not need to be localised at a particular post (in our simulations we change the contact angle along all the channel); it is sufficient that it covers all the posts where filling is (re)-started by the applied potential. This avoids the challenge of miniaturising the electrodes.

\begin{figure}[!h]
\includegraphics[angle=0,scale=0.25]{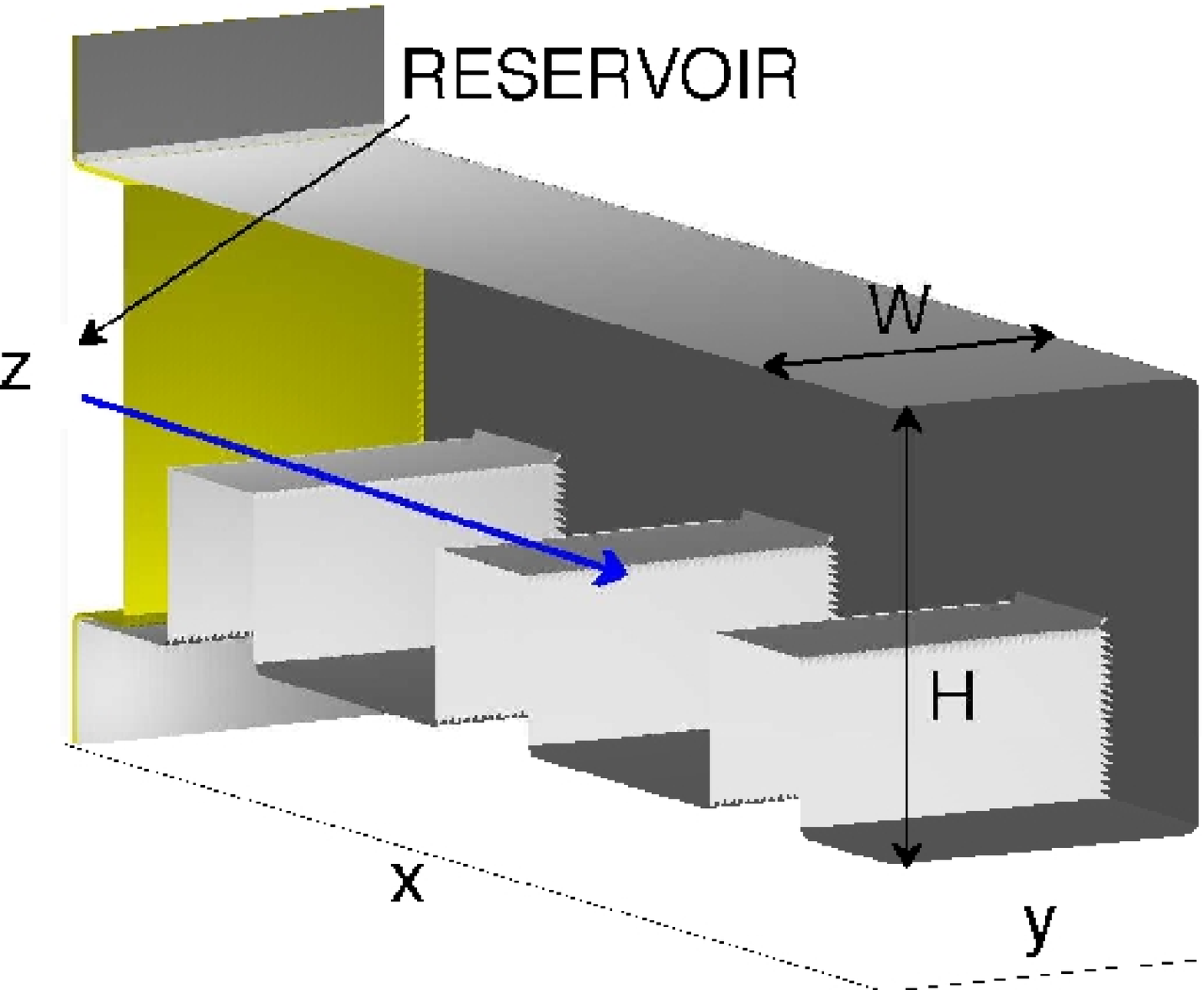} (a) \\ \vspace{0.7cm}
\includegraphics[angle=0,scale=0.25]{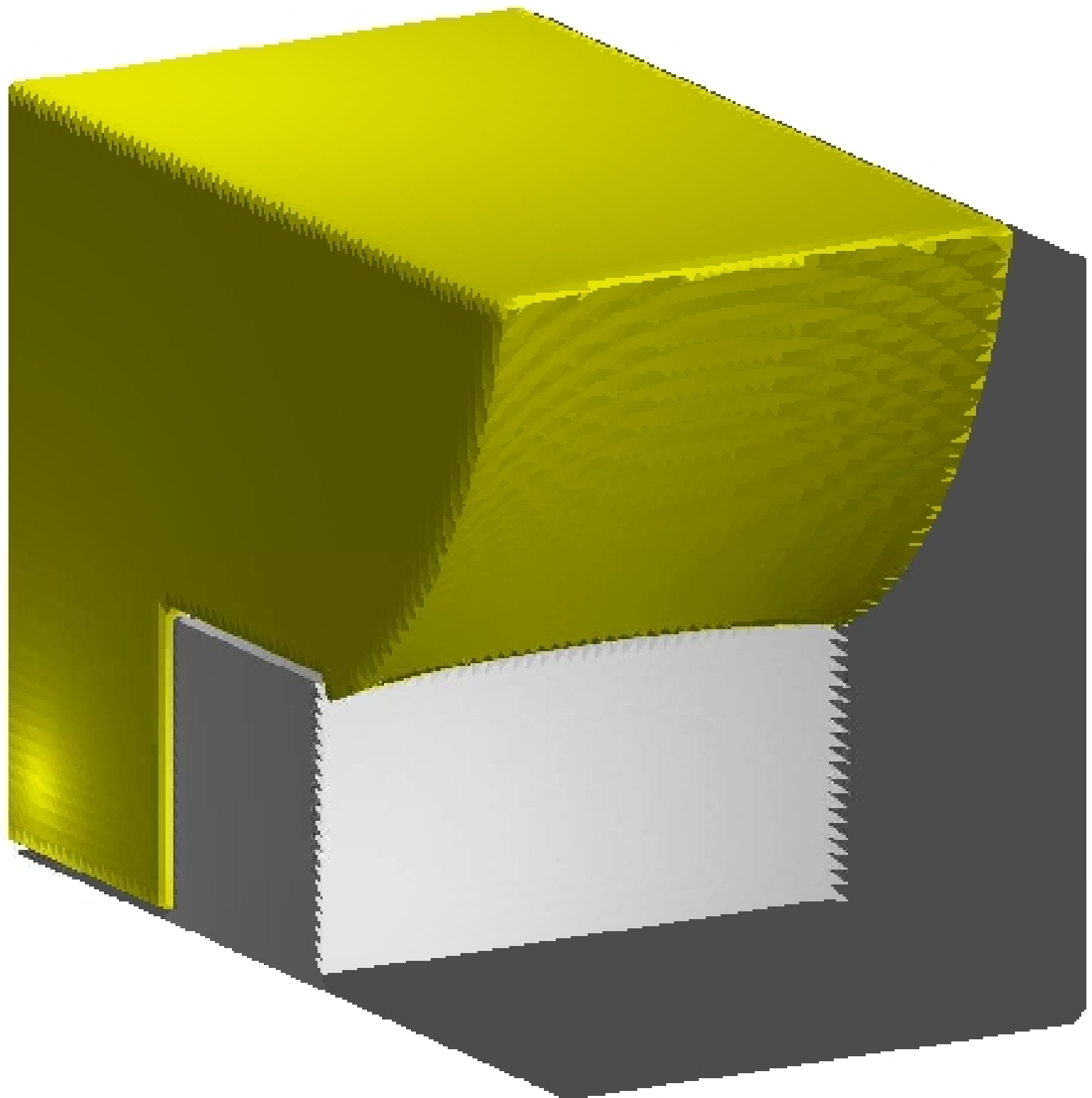} (b) \\ \vspace{0.7cm}
\includegraphics[angle=-90,scale=0.25]{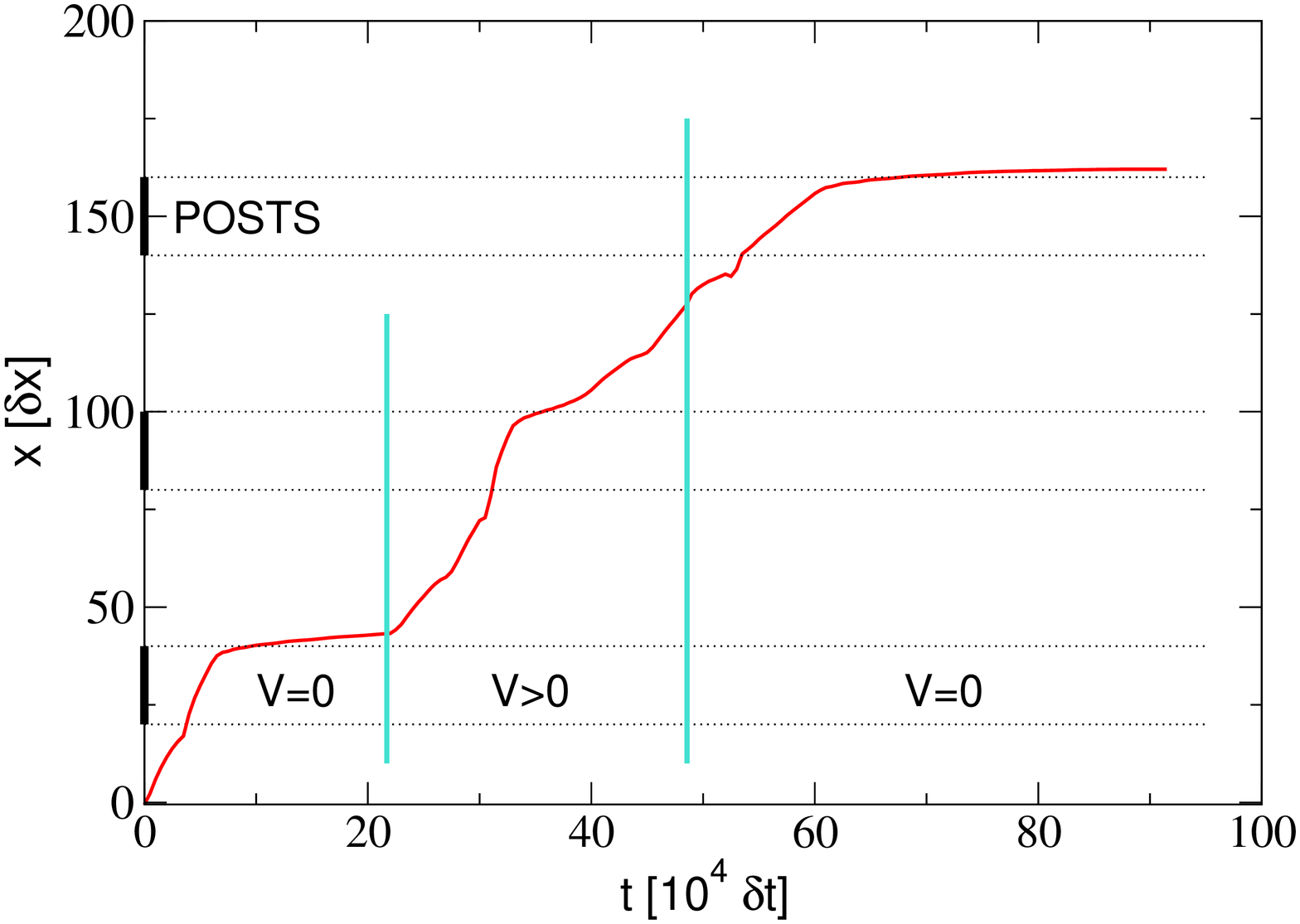} (c)
\caption{
(a) Channel geometry: the fluid flows along the $x$ direction. (b) Typical shape of the front 
when it is pinned at a ridge. For clarity the top and a side wall are omitted. (c)  Position 
of the filling fluid  versus time. In these simulations $\theeq=70^\circ$, $\thev=55^\circ$ 
and $\thetr \sim 65^\circ$. 
}\label{groove}
\end{figure}
We next consider a second channel geometry, for comparison, and to illustrate the robustness of the mechanism. This is shown in Fig.~\ref{groove}{\em a}. The channel has height $H=50$ and width $W=40$ and side walls are now explicitly included. Only the bottom wall of the channel is patterned, by ridges of height and width 20 and spacing 40 which run all across its width. The aspect ratio of this channel, and the fact that only one wall is patterned, may make it more accessible experimentally. 

Running simulations to locate the threshold value
of the equilibrium contact angle gives a value of 
$\thetr$ between  $65^\circ$ and $70^\circ$. A pinned configuration with
$\theeq=70^\circ$ is reported in Fig.\ \ref{groove}{\em b}. 
(For a two dimensional geometry  with no side walls $\thetr=45^\circ$.
Thus, as expected, the hydrophilic side walls significantly increase 
ease of filling.)

Results of simulations for the grooved channel are shown in Fig.\ \ref{groove}{\em c}. The simulation procedure was analogous to that used for the channel with posts, but with the contact angle cycling from $\theeq=70^\circ$ at $t=0$ to $\thev=55^\circ$ at $t = 2.2 \cdot 10^5 \, \delta t$ and back to $70^\circ$ at $t = 4.9 \cdot 10^5 \, \delta t$. The pattern of filling is very similar; as before the interface is halted at the first and third posts (compare Figs.~\ref{fig1}{\em c} and Fig.\ \ref{groove}{\em c}).

If the electrowetting potential is applied to the top wall only, the front is still depinned but touches the subsequent post before the bottom wall is completely wet. This leads to the formation of bubbles of air trapped between the posts. This could be avoided if necessary by choosing smaller ridges or more widely separated posts.
An estimate of the geometry necessary to avoid bubble formation follows from observing that the advancing front makes an angle $\theeq$ with the dry face of the ridge as it depins. Therefore to avoid the front touching the subsequent ridge without having completely wet the previous one, hence trapping a bubble, it is sufficient that the ratio of the height of the posts to the distance between them is less than $\tan(\pi/2-\theeq)$. To check this we ran faster, two-dimensional simulations for the grooved geometry, similar to that reported in Fig.\ \ref{groove}{\em a} but without lateral walls and with the electrowetting potential applied just to the top wall. Keeping the equilibrium contact angle on the patterned substrate fixed at 60$^\circ$, we switched the top wall between 60$^\circ$ and 25$^\circ$ reproducing a filling profile similar to that reported in Fig.\ \ref{groove}{\em c} and with no air entrainment.

In this paper we have demonstrated that it is possible to use mesoscale simulations to study the behavior of an interface in a patterned microchannel. We have shown that surface patterning, together with electrowetting, can be used to control how the interface advances along the microchannel. We do not expect the valve mechanism to depend qualitatively on the size of the channel as long as the channel is sufficiently small that gravitational effects can be ignored. The obstacles must be at least a few microns in size, so that the fluid cannot overcome the pinning due to thermal fluctuations, and the advancing fluid must be non-volatile or it will move past the obstacles by evaporating and then condensing.

However there is a strong quantitative dependence of $\thetr$ on the details of the patterning allowing this to be adjusted to lie between the available $\theeq$ and $\thev$. Moreover, although we have considered the quasistatic case, the threshold angle will increase for an interface with finite velocity or for pressure driven flow. 

We hope that this paper will motivate experiments on patterned microchannels, and that simulations such as those presented here will be useful in determining suitable channel geometries for a given application and materials. Similar modelling approaches could also be used to predict when an interface depins from the end of a  constriction in a microchannel or a micronozzle due to an applied pressure \cite{DGLSK-99}.

\end{document}